\documentclass[12pt,preprint]{aastex}
\usepackage{emulateapj5}
\usepackage{epsf}
\begin{document}
\shorttitle{Old-Population High Velocity Stars}
\shortauthors{Kollmeier \& Gould}
\newcommand{\msun}{M_{\odot}}
\newcommand{\kms}{\, {\rm km\, s}^{-1}}
\newcommand{\cm}{\, {\rm cm}}
\newcommand{\gm}{\, {\rm g}}
\newcommand{\erg}{\, {\rm erg}}
\newcommand{\kel}{\, {\rm K}}
\newcommand{\kpc}{\, {\rm kpc}}
\newcommand{\mpc}{\, {\rm Mpc}}
\newcommand{\seg}{\, {\rm s}}
\newcommand{\kev}{\, {\rm keV}}
\newcommand{\hz}{\, {\rm Hz}}
\newcommand{\etal}{et al.\ }
\newcommand{\yr}{\, {\rm yr}}
\newcommand{\mpyr}{{\rm mas}\, {\rm yr}^{-1}}

\newcommand{\gyr}{\, {\rm Gyr}}
\newcommand{\eq}{eq.\ }
\def\arcsec{''\hskip-3pt .}
\def\prine{{}} 
\def\ch{\bf }
\title{Where are the Old-Population High Velocity Stars?}
\author{Juna A. Kollmeier\altaffilmark{1,2} and Andrew Gould \altaffilmark{3}}

\altaffiltext{1}{Observatories of the Carnegie Institution of Washington, 813 Santa Barbara Street, Pasadena, CA 91101}
\altaffiltext{2}{Hubble Fellow, Carnegie-Princeton Fellow}
\altaffiltext{3}{Dept. of Astronomy, The Ohio State University,
140 W. 18th Ave, Columbus, OH 43210}

\begin{abstract}

To date, all of the reported high velocity stars (HVSs), which are
believed to be ejected from the Galactic center, are blue and
therefore almost certainly young.  Old-population HVSs could be much
more numerous than the young ones that have been discovered, but still
have escaped detection because they are hidden in a much denser
background of Galactic halo stars.  Discovery of these stars would
shed light on star formation at the Galactic center, the mechanism by
which they are ejected from it, and, if they prove numerous, enable
detailed studies of the structure of the dark halo.  We analyze the
problem of finding these stars and show that the search should be
concentrated around the main-sequence turnoff $(0.3<g-i<1.1)$ at
relatively faint magnitudes $(19.5<g<21.5)$.  If the ratio of turnoff
stars to B stars is the same for HVSs as it is in the local disk, such
a search would yield about 1 old-population HVS per $10\,{\rm deg}^2$.
A telescope similar to the Sloan 2.5m could search about $20\,{\rm
deg}^2$ per night, implying that in short order such a population,
should it exist, would show up in interesting numbers.
\end{abstract}

\section{Introduction
\label{sec:intro}}

Hypervelocity stars (HVSs---stars with velocities in excess of the
Galactic escape speed) have come a long way since \citet{hills88}
predicted their existence.  It is now appreciated that, beyond being a
dynamical curiosity, these stars are useful probes of multi-scale
Galactic phenomena.  Their frequency, spectral properties, and
distribution provide important constraints on the character of star
formation in the Galactic Center (GC) as well as the stellar ejection
mechanism itself.  Furthermore, in sufficient numbers these objects
are unique dynamical tracers of the shape of the Milky Way's dark
matter halo---a critical quantity in understanding how the Galaxy fits
into the overall picture of hierarchical structure formation
\citep{gnedin05}.

Owing to their rapid Galactic exit times and low predicted ejection
rates for plausible dynamical mechanisms (e.g., \citealt{yu03}), these
stars are relatively rare.  The first HVS discovery was a
serendipitous byproduct of a kinematic survey of blue horizontal
branch (BHB) stars \citep{brown05}.  Through spectroscopic follow-up
of 36 faint ($19.75 < g < 20.5$) color selected $(0.8 < u - g <
1.5)\cap(-0.3 < g - r < 0.0 )$ BHB stars from the Sloan Digital Sky
Survey (SDSS) First Data Release, these authors discovered a
$6\,\sigma$ radial-velocity outlier at $709\,\kms$ with $g = 19.81\pm
0.02$ and dereddened colors $(u - g)_0= 1.04\pm 0.09$ and $(g - r)_0=
-0.30\pm 0.03$.  This star was subsequently determined to be a
pulsating B-type main sequence star \citep{fuentes06}. Shortly after
this discovery, two additional HVSs were found, also within surveys
designed for the selection of early-type stars.  During their survey
for subluminous B stars (sdB) \citet{edelman05} discovered an $8\msun$
B star with radial velocity of $563\,\kms$ located at $\sim
60\,\kpc$ from the Galactic center, potentially ejected from the LMC.
\citet{hirsch05} followed up star US 708 as part of a survey of
$\sim 100$ subluminous O stars selected from SDSS having colors $(u -
g) < 0.2$ and $(g - r) < 0.1$, finding it to be a Helium rich
subluminous O star (HesdO) traveling at $\sim 720\,\kms$ at $25\,\kpc$ from
the GC. After these initial discoveries, in which HVSs were
contaminants in surveys for other blue stars, \citet{brown06}
undertook the first {\it targeted} survey for HVSs, in which
candidates were selected to be relatively faint, ($17.5 < g < 18.5$),
with B-star colors, $[-0.42<(g-r)<-0.27] \cap [1.33 < (u - g) - 2.67(g
-r) < 2.0 ]$.  This search strategy has two key components: maximizing
both the volume covered and the contrast with normal halo stars.  The
faint magnitudes achieve the first aim, while both the color and
magnitude selection contribute to the second.  Contrast is improved at
faint magnitudes (large distances) because the HVS density should drop
off as $R^{-2}$, where $R$ is Galactocentric distance, while the
normal halo stars drop off faster than $R^{-3}$. It is improved
at blue colors because the only blue halo stars are BHB stars,
which have short lifetimes and so low density, and white dwarfs,
which are even rarer.  Midway through their survey, \citet{brown06}
have found 4 probable HVSs out of 192 candidates culled from
$3000\,\rm deg^2$, i.e., a density of $(1/750)\,\deg^{-2}$.

All reported HVSs are blue, which simply reflects the fact that, after
the first three serendipitous discoveries, the searches were conducted
among blue stars. One can imagine extending previous work 
in several possible directions in parameter space, for example
by searching for the A stars emerging from the same
underlying population as the already discovered B stars.  If the HVSs
are typical of bulge stars, however, there should be many {\it old}
ones.  Due to the high background of unevolved halo stars, no one has
yet undertaken the daunting (and seemingly hopeless) task of a
comprehensive survey for this {\it late-type} population of HVSs.

 However, determining whether this population exists and in what
proportion would be of great interest.  In particular the ratio of old
to young HVSs would place important constraints on the stellar
ejection mechanism itself.  It is possible that the distribution of
HVS ages reflects the distribution in the stellar cusp near the GC.
Models that suggest HVS ejections are due to a short burst of
scatterings from an intermediate mass black hole (IMBH) that falls to
the GC by dynamical friction and disrupts the stellar cusp there
\citep{baumgardt06} would be directly tested by knowledge of the HVS
age distribution.  With high-precision proper motion measurements of a
sufficient number of HVSs, one could measure not only the Galactic
potential but also the times of ejection for individual stars
\citep{gnedin05}.  Regardless of the relative number of late- to
early-type HVSs, measurement of this ratio would be interesting.  If
the ratio proved small, this observational fact could constrain
pictures in which the young stars near Sgr A* are brought to the
GC in clusters along with an IMBH as suggested by
\citet{hansen03}.  It is also possible that late-type HVSs vastly
outnumber early-type HVSs, but the difficulty in finding this
population has prematurely biased our view of it.  Should this be the
case, such stars would be vital probes of Galactic structure.

In brief, the true age distribution of HVSs is simply unknown.
We therefore turn to the problem of how to find the old population.
Clearly, it is not practical at the current time to attempt spectroscopy 
of all halo stars.  In \S~\ref{sec:needle} we analyze the problem
of developing optimal selection criteria to search for them.
Then, in \S~\ref{sec:discuss} we comment on the prospects for detecting
HVSs in future surveys.

\section{Needles in a Haystack
\label{sec:needle}}

Old-population HVSs must be much less common than even the relatively 
low-density population of halo stars.  Otherwise, they would have been 
discovered in spectroscopic surveys of high proper-motion stars.
Hence, finding such stars against the much more numerous background
of halo stars will require a well thought-out search strategy.

If HVSs are ejected isotropically from the Galactic center at a rate
$\Gamma$, then their density at Galactocentric distance $R$ is 
\begin{equation}
\rho(R) = {\Gamma\langle[v(R)]^{-1}\rangle\over 4\pi R^2},
\label{eqn:rho}
\end{equation}
where $v(R)$ is the velocity of the ejected star as a function of $R$
and where the brackets indicate averaging over the inverse velocities.
Note that ``$\Gamma$'' could refer to the HVS population as a whole
or to any subclass of stars within it.

\subsection{Zeroth Order Analysis
\label{sec:zero}}

To facilitate the exposition, we make a set of simplifying assumptions.
Taken together, these lead to a toy model that, while not realistic, does 
yield a useful starting point for understanding the problem of finding HVSs.  
In the next section we will sequentially relax these assumptions, allowing 
the features of the real problem to come into focus.

First, we assume that HVSs do not decelerate as they leave the Galaxy.
Equation (\ref{eqn:rho}) then simplifies to $\rho(R)\propto R^{-2}$.
Second, we assume that the physical density of halo stars also scales
$\rho_{\rm halo}\propto R^{-2}$.  Then, under this very unrealistic
assumption, the ratio of HVSs to halo stars would have the same
constant value at any position in the Galaxy.  Finally, we assume that
the color-magnitude relation of the old-population HVSs is identical
to that of halo stars.  Together, these assumptions would imply that
the ratio of HVSs to halo stars is exactly the same for candidates
selected at any color and apparent magnitude and in any direction,
provided only that the selection criteria ensured that disk and
thick-disk stars were effectively excluded.

\subsection{First Order Analysis
\label{sec:first}}

As each of the three assumptions is relaxed, the fraction of HVSs
increases with increasing $R$.  First, from equation (\ref{eqn:rho}),
progressive deceleration makes the density of HVSs fall more slowly
than $R^{-2}$.  Second, the density of halo stars falls much
more quickly than $R^{-2}$.  Locally halo stars fall roughly as
$R^{-3.2}$ (e.g.\ \citealt{gould98}), and the relation steepens
further out.  Third, stars near the Galactic center are generally more
metal rich than halo stars, which implies that they are more luminous
on the upper main sequence.  As we will show below, upper
main-sequence and turnoff stars dominate the HVS discovery potential.
The fact that HVSs are more luminous implies that they lie at farther
distances at fixed magnitude.  Hence, they cover a larger range of
distance over a fixed magnitude range than the corresponding halo
stars.  This enhances their density as a function of apparent
magnitude.

Thus, other factors being equal, one should try to search for
HVSs as far from the Galactic center as possible.  Within a given
field, ``other factors'' are obviously not equal because it is
easier to search among bright than faint stars.  So this issue will
require additional analysis.  However, we are at least driven to
the conclusion that the search for HVSs will be easiest toward
high-latitude, Galactic-anticenter fields: high-latitude to avoid
contamination from disk stars, and anticenter to reach the maximum
$R$ at fixed apparent magnitude.

\subsection{Characteristics of Background
\label{sec:background}}

\begin{figure}
\plotone{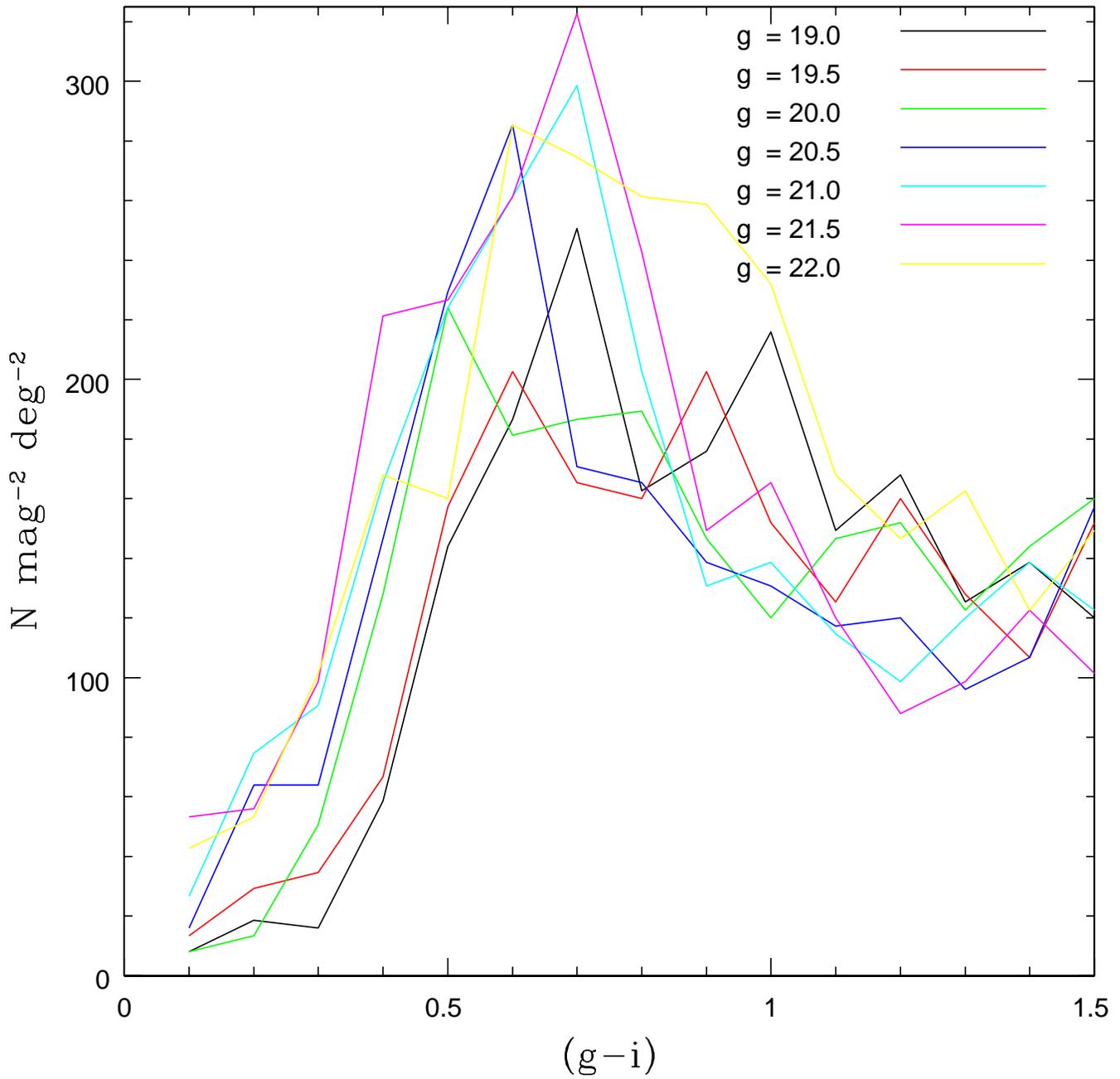}
\caption{Density ($\rm mag^{-2}\,deg^{-2}$) of stars in the color
magnitude diagram of the SDSS high-latitude anti-center field toward
$(l,b)=(196,-40)$. Curves correspond
to constant-magnitude bins at half-magnitude intervals from $g=19$ to
$g=22$.  To a first approximation, all the curves are the same.
Moreover, they are roughly flat redward of the turnoff (at $g-i\sim
0.65$) and even on the turnoff they are only about double the plateau
value.  }
\label{fig:cmd}
\end{figure}

Because the selection criteria for HVS candidates can only be color 
and apparent magnitude, we must begin by analyzing the background
in terms of these two variables.  We adopt a purely empirical approach,
tabulating the density of stars toward a SDSS
field centered at approximately RA=4 hr , Dec=$-6^\circ$
($l=196$, $b=-40$).  Figure \ref{fig:cmd} shows the stellar density
as a function of $g-i$ color for 7 different magnitude bins centered
on $g=19.0, 19.5.,\ldots 22.0$.
The key point is that in this magnitude range and in the region from the 
turnoff redward, this density varies by less than a factor 4 altogether
and only by a factor 2 in the basic trend with color.  
Moreover the color profiles at
each $g$ magnitude are approximately the same.
These characteristics imply that the
background does not play a crucial role in devising selection
procedures for HVS candidates as it would have were the curves
clearly separated: rather color/magnitude selection
must be based primarily on maximizing the total number of HVSs
and minimizing the amount of observing time required to identify
them as HVSs.  We return to the issue of turnoff vs. giant and lower MS stars in \S~\ref{sec:color}.

\subsection{Color Selection
\label{sec:color}}

Under certain simplifying assumptions, the color/magnitude selection
actually factors into separate selections in color and magnitude.
We first introduce and motivate these assumptions and later evaluate
how their relaxation would impact our conclusions.

First, we assume that deceleration is negligible so that, as mentioned
in \S~\ref{sec:zero}, $\rho\propto R^{-2}$.  In fact, for an isothermal
sphere of circular velocity $v_{\rm circ}=220\,\kms$, the velocity-squared
falls by $\Delta v^2 = 2v_{\rm circ}^2\ln(R_2/R_1)$ between $R_1$
and $R_2$.  For example, a star traveling at $800\,\kms$ at $15\, \kpc$
will slow by 15\% to $675\,\kms$ at $100\, \kpc$.  This is not completely
negligible but it is modest compared to other factors in the problem.
Under this assumption, the number of HVSs of a fixed absolute magnitude (and 
so by assumption fixed color) and a narrow range of
apparent magnitudes $\Delta g$, is
\begin{equation}
N = {\ln 10\over 5}\,{\Gamma\langle v^{-1}\rangle\over 4 \pi}\,
\Omega\Delta g
{r^3\over R^2}
\label{eqn:ntrue}
\end{equation}
where $\Omega$ is the angular size of the field and $r$ is the distance 
from the observer to a star at the center of the magnitude bin.

Second, we assume that $R=r$, which reduces the last term in equation
(\ref{eqn:ntrue}) from $r^3/R^2\rightarrow r$.  That is, 
$N_{\rm naive}\propto r$.
The ratio of this naive estimate to the true number is
\begin{equation}
{N_{\rm naive}\over N} = 1 - 2\cos l\cos b{R_0\over r} + {R_0^2\over r^2}
\rightarrow 1 + 1.47{R_0\over r} + {R_0^2\over r^2},
\label{eqn:nnaive}
\end{equation}
where $R_0=8\,\kpc$ is the solar Galactocentric radius.
Clearly this correction can be fairly large, so we will have to 
carefully assess its impact after the selection criteria are derived.

Because radial velocity (RV) measurements are most efficiently
carried out in the $g$-band part of the spectrum, RV precision for
a fixed exposure time is basically a function of $g$ magnitude.
This is not exactly true because the metal lines, from which these
determinations are primarily derived for FGK stars, are stronger at
lower temperatures.  However, this is a modest correction, which we
will ignore for the moment but to which we will return below.

Consider now an ensemble of HVSs, drawn randomly from a common
old-star isochrone and ejected isotropically and stochastically
from the Galactic center.  We now select stars at a fixed $g$-magnitude
(or rather in a narrow interval $\Delta g$ centered at fixed $g$),
which have a variety of absolute magnitudes $M_g$ and so (through
the color-magnitude relation of the isochrone), a variety of $g-i$ colors
(which are what we actually observe).  The stars at $M_g$ will
be seen over a range of distance 
$\Delta r = (\ln 10/5)10^{0.2(g-M_g+5)}\Delta g\,$pc, i.e.
$\Delta r \propto 10^{-0.2\,M_g}$.  Under the above two assumptions,
the relative number of such stars in the sample will be
\begin{equation}
N_{det}(M_g) \propto 10^{-0.2\,M_g} \Phi(M_g), 
\label{eqn:lumdep}
\end{equation}
where $\Phi(M_g)$ is the
fraction of stars from the isochrone in the $M_g$ bin.  Note, in
particular that this relative number does not depend on $g$,
the apparent magnitude at which they are selected.

\begin{figure}
\plotone{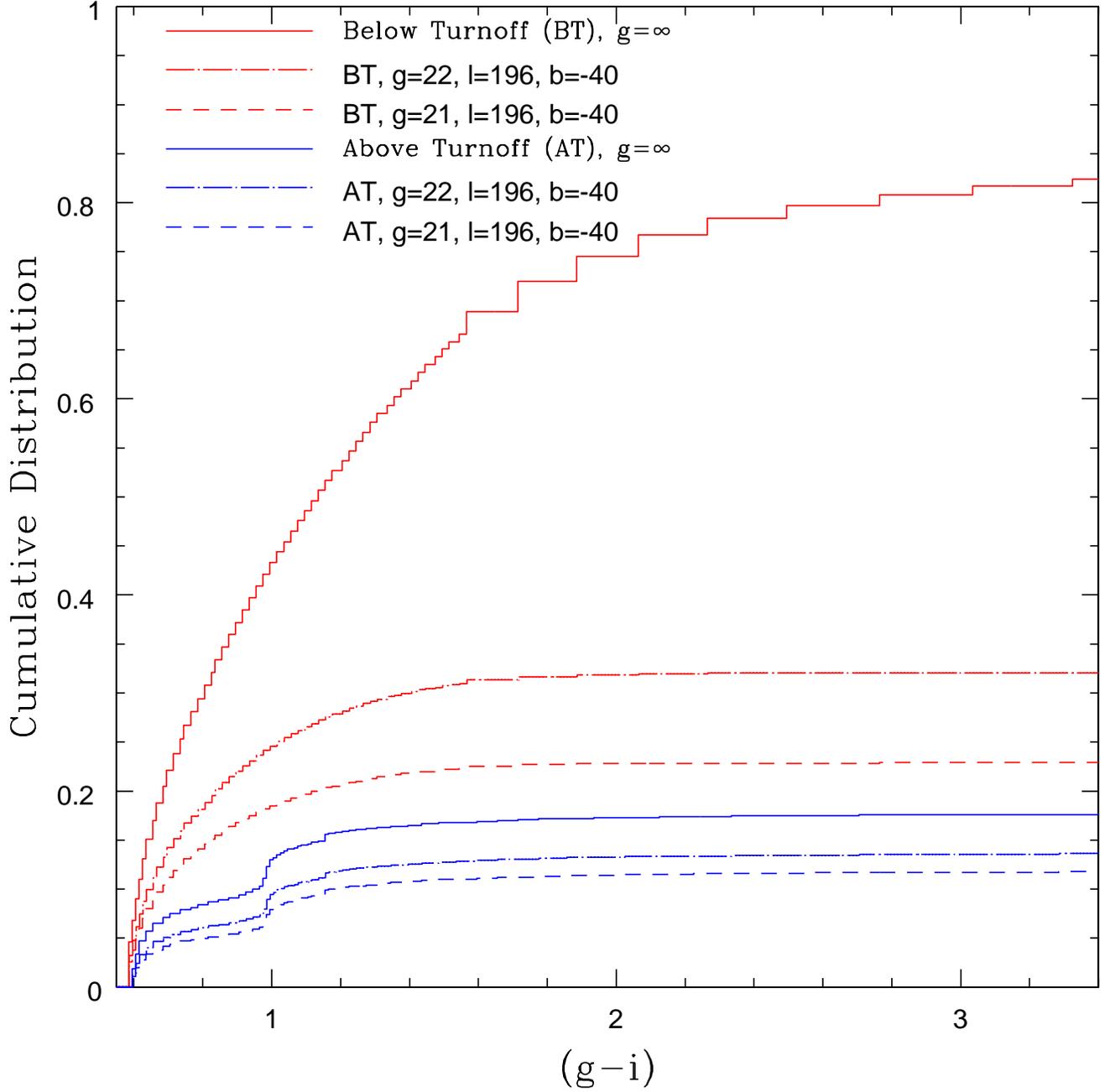}
\caption{Cumulative distribution of expected high-velocity-star detections
of apparent magnitude $g$
as a function of $g-i$ color, under the assumption that all stars
in the underlying old-population are equally likely to be ejected from
the Galactic center.  The red curves represent stars below the main-sequence
turnoff while the blue curves represent stars above the turnoff.  
At very faint magnitudes, the result is independent of magnitude and
is shown by the two {\it solid curves}, whose total is arbitrarily
normalized to unity.  At realistic magnitudes, $g=22$ ({\it dot-dashed}) or
$g=21$ ({\it dashed}), the detections are
somewhat suppressed by eq.~(\ref{eqn:nnaive}), particularly for the
below-turnoff branch.  For these curves, the great majority of expected
detections are at $g-i<1.1$.
}
\label{fig:cumulative}
\end{figure}

Figure \ref{fig:cumulative} shows the cumulative distribution of these
relative numbers as a function of color (the observed quantity) for an
isochrone of solar metallicity and age of 10 Gyr \citep{yale}.  Since
these are given in the Johnson/Cousins system, we convert to SDSS
bands using the transformations given on the SDSS web site
\citep{lupton05}.  When the isochrone is viewed as a ``function'' of
color, it is double-valued.  To illustrate the role of the two
branches, we plot their cumulative distributions separately, although
of course these could not be distinguished from color/magnitude data
alone.  The solid curves illustrate the result under the assumption
that $r=R$, i.e., effectively that $g=\infty$. The dashed and
dot-dashed curves are for the more realistic cases of $g=22$ and
$g=21$ in the direction $(l,b)=(196^\circ ,-40^\circ)$.

There are several important features of this diagram.  First, and by
far the most important, the great majority of potential sensitivity to
old-population HVSs comes from stars with $g-i$ colors within 0.5
magnitudes of the turnoff, i.e., with $g-i<1.1$.  This is already
basically true for the naive ``$g=\infty$'' case, but strictly applies
in realistic cases, $g\la 22$.  By contrast, both the M and late-K dwarfs
on the lower branch and the M and late-K giants on the upper branch,
contribute very little, the former because they are so close and the
latter because they are so rare.  Note that the dominance of turnoff
stars is a specific result of the $L^{1/2}$ luminosity-dependence in
equation~(\ref{eqn:lumdep}).  If $N\propto L^{3/2}$ (as in a
magnitude-limited sample of uniform-density population) then giants
would dominate.  If $N\propto L^0$ (as in a magnitude-limited sample
of an $r^{-3}$ halo-star-like population) then dwarfs would dominate.
Second, the lower branch contributes a bit more than double the upper
branch for realistic cases.  That is, the sample is dominated by stars
just below the turnoff, with a significant, though clearly secondary,
contribution from stars just above the turnoff.  This implies that the
validity of our approximations is basically determined by how well
they hold up at the turnoff.  Third, the small contribution from
late-type giants obviates another potential complication.  Depending
on the precise form of the ejection mechanisms, it is possible that
giant-star ejection is suppressed relative to smaller stars.  For
example, some or all of the ejections might take place from disruption
of relatively tight binaries that are too close to permit giant-star
survival.  Had Figure~\ref{fig:cumulative} implied giants dominated
the HVS distribution, this would lead to significant uncertainty.
However, the small contribution of giants, particularly late-type
giants, implies that any such suppression would also have small
impact.  The one exception to this is the clump giants, which are not
included in the Yale Isochrones and hence are not represented in this
figure.  They would contribute a small ``bump'' in the ``above
turnoff'' curve, similar in amplitude to the bump at $g-i\sim 1$ that
is actually seen in this curve, which is due to first-ascent giants.
With radii ten times solar, these stars are themselves relatively
small.  However, with ages of only $100\,$Myr, they are younger than
the transport time to their current location, roughly $200\,$Myr at
$g=21$ and velocity $v=700\,\kms$.  Hence, the progenitors of these
stars would have had to have been ejected when they were very
distended.  In any event, they are not included in the figure.
Finally, the slightly greater RV precision (at fixed $g$ and fixed
exposure time) of cooler stars also has negligible impact, again
because of the small contribution of these stars.

 From this analysis, we conclude that at fixed magnitude, selection
should stretch from the turnoff 
and proceed redward to $g-i=1.1$.  In practice, the old-population
HVSs will not come from a single isochrone, but from a superposition
of many isochrones with a variety of ages and metallicities.  However,
all of these are qualitatively similar, with just slightly varying
turnoff colors.  Indeed, we investigated a 5 Gyr isochrone and
found results qualitatively similar to those shown in 
Figure \ref{fig:cumulative}.
The important practical point is just to sample
the field stars beginning blue enough to cover all such turnoffs.
The cost of moving the blue boundary further blueward by $\Delta (g-i)=0.3$
is quite small, since this region of the observed field-star color-magnitude
diagram has few stars.  We therefore advocate color selection $0.3<g-i<1.1$.

\subsection{Magnitude Selection
\label{sec:magnitude}}

Observations of a fixed exposure time can potentially measure RVs
to a given precision down to a certain apparent-magnitude limit $g$.
Once this is established, one could in principle measure RVs for all
stars within this limit, or (if fibers/slits were scarce) only those
within 2 magnitudes of the limit, which would contain $>60\%$ of all the
HVSs within the magnitude limit.  
The following arguments apply equally to either strategy.

Let us first suppose that ``downtime'' (for slewing, readout, and
changing slit masks or fiber positions) is negligible compared to the
exposure time.  Let us compare two observation strategies, the first
with a single field exposed for time $\Delta t$ and the second with
two fields each exposed for $\Delta t/2$.  Let us initially assume
that the magnitude limit is above sky in both cases.  Then the flux
limit will be a factor 2 larger in the second case to maintain the
same signal-to-noise ratio (S/N).  The maximum observable distance
will therefore be reduced by $2^{-1/2}$, which will decrease the
number of HVSs detected in each field by the same factor.  However,
since there are twice as many fields, the total number of detected
HVSs will increase by $2^{1/2}$.  Hence it is always better to go to
shorter exposures of more fields.  If both limits are below sky, then
the flux limit increases by $2^{1/2}$, so the distance limit decreases
by $2^{-1/4}$, and the number of HVSs from both fields increases by
$2^{3/4}$.  That is, the same argument applies even more strongly.

Now consider the opposite limit, in which the exposure time is
negligible compared to the downtime.  Shortening the exposure time
then still increases the flux limit and so reduces the number of HVSs
detected by $2^{-1/2}$ but in this case there is no compensating
increase in the number of fields covered.  Comparing the two cases, it
is clear that the exposure times should be set approximately equal to
the downtime.  It can be shown that the optimal exposure time is
exactly equal to the downtime if the magnitude limit is above sky and
equal to 1/3 of the downtime if it is below the sky.

This argument somewhat overstates the case: it would be strictly valid
if the cumulative distributions illustrated in
Figure~\ref{fig:cumulative} were identical for the limiting magnitudes
corresponding to the two different exposure times.  These curves are
nearly identical for the upper branch, but less so for the lower
branch.  However, the argument remains qualitatively valid, the
correction being toward exposures that are somewhat longer than the
downtime.

\subsection{Observing Strategy: General Considerations
\label{sec:generalities}}

Before analyzing the characteristics of specific spectrographs, there
are two general points to consider.  First, from Figure \ref{fig:cmd},
the stellar density in our recommended color range, $0.3<g-i<1.1$, is
about $150\,\rm mag^{-1}\,deg^{-2}$.  Hence, if one is to cover 2 or 3
magnitudes in $g$, this requires monitoring 300--500 stars per square
degree.  Note that in another direction, $(l,b)=(275^\circ,62^\circ)$,
we find stellar densities in this color-mag range that are about 2.5
times higher, confirming that it is substantially easier (in terms of
the sheer number of background contaminants) to search for
old-population HVS in high-latitude anticenter fields.

Second, the RV precision requirements to distinguish HVSs from halo stars
are not very severe: $\sigma=50\,\kms$ would be quite adequate.  Typical
HVSs have RVs $v_r\sim 700\,\kms$, well separated from halo stars at
$|v_r|\sim O(170\,\kms)$.  There are, of course, halo stars moving
closer to the escape velocity, but these are relatively rare, and the
chance that one of these would further upscatter by $\sim 5\,\sigma$
into the HVS range is quite small.  This precision is not difficult
to achieve even in very noisy spectra, particularly on FGK stars,
which have many spectral features.

\subsection{Specific Evaluations
\label{sec:specifics}}

 From this point forward, concrete development of an observing strategy 
obviously depends
on the detailed characteristics of the multi-object spectrograph, which
cannot be treated completely generally.
However, to give some broad guidance and to help understand the sensitivity
of the search under realistic conditions, we consider two specific
multi-object spectrographs with radically different characteristics.

First, we consider the resolution ${\cal R}= 2000$ SDSS spectrograph, which has 
640 $3''$ diameter fibers spanning a $7\,\rm deg^2$ field on a 2.5 m telescope.
The fiber plug-plates require about 10 minutes to change.  In principle,
one should consider the time required for repointing the telescope,
but this will be relatively infrequent because even in the anticenter fields,
there are about 2000 viable targets (for a 2-magnitude interval) and only
640 fibers, so there should be about three plug-plates per pointing.
If we strictly applied the ``exposure time equals downtime'' 
rule, the resulting 10 minute
exposure would yield per pixel S/N=10 
at about $g=19.5$. Given the $3''$ fibers, this
is about 1 mag below dark sky.  In practice, it is probably impractical
to change plug plates so frequently, so 45 minute exposures (in keeping
with current SDSS practice) would appear more realistic.  Taking account
of sky, this yields a per pixel S/N=8 at $g=20.5$ and S/N=4 at $g=21.5$.  
This latter is probably adequate for $\sigma=50\,\kms$ measurements, but this
should be tested directly.  Hence, in one 9-hour night, the SDSS
telescope could cover a total of $21\,\rm deg^2$ over $19.5<g<21.5$ and
$0.3<g-i<1.1$, i.e., 3 45-minute exposures (1 for each of 3 plug plates)
on each of 3 fields.  One additional wrinkle should be noted.  The
640 fibers are divided into 320 red and 320 blue channels, with
the dividing line at 6000$\,$\AA .  The blue-channel fibers are
more sensitive to RV because of the greater density of lines.
However, this can easily be compensated by applying the red-channel
fibers to the brighter stars.  Being systematically 1 magnitude
brighter (and recalling that the target stars are mostly below sky),
these would have about 2.5 times higher S/N per pixel.

Second, we consider the ${\cal R}=20000$ IMACS F/4 spectrograph, which
accommodates up to 1000 slits spanning a $0.067\,\rm deg^2$ field on
the Magellan 6.5 m telescope.  Changing slit masks requires about 15
minutes, but in fact this is not the relevant scale of ``downtime''
because each field has only of order 20--30 available targets, far
fewer than the available slits.  Rather, each mask could be cut to
serve of order 10 fields (with a total of 200-300 targets).  Hence,
the downtime is primarily set by the time required to acquire a new
field (without changing the mask).  This is roughly 5 minutes, which
by the guideline derived in \S~\ref{sec:magnitude} would indicate an
exposure time also of 5 minutes.  Taking account of sky noise and 
assuming $4\,e^{-}$ read noise, this leads to an estimate of per
pixel S/N=1.2 at $g=21.5$.  To evaluate the utility of such
signal levels, we construct synthetic spectra, add noise, and then
fit the results to a (4-parameter) quadratic polynomial plus a template
spectrum, offset by various velocities from the constructed spectrum.
We find that this S/N is sufficient for an
accurate RV measurement, provided that at least $80\,$\AA\ are sampled
centered on $\lambda=5175\,$\AA .  In fact, the formal error in the
measurement is less than $10\,\kms$, so that it would appear that even
lower S/N would be tolerable.  However, we find if the S/N is further
reduced, that while the width of the correlation peak does not
increase dramatically, multiple minima (each quite narrow) begin to
appear, undermining the measurement.  Similarly, if the wavelength
coverage is reduced at constant S/N, then multiple minima also appear.
In any event, if appropriate blocking filters permit this $80\,$\AA\
(800 pixel) window or larger, 
then the field can be reliably probed to $g=21.5$.
Of order 50 fields could be searched during a 9 hour night, covering
about $3.3\,\rm deg^2$.

Thus, while these two spectrographs differ in aperture, resolution,
and field size by factors of 7, 10, and 100 respectively, they
are capable of broadly similar searches for HVSs.  We conclude
that it is feasible to conduct the search over tens of square degrees
on a variety of telescopes without exorbitant effort.

\section{Discussion
\label{sec:discuss}}

How likely is it that such a survey of several tens of square degrees
will detect old-population HVSs?  At one level, as emphasized in 
\S~\ref{sec:intro}, we have no idea:  based on what we know now,
old-population HVSs could equally well be very common or non-existent.
However, in the absence of any hard information, we might guess
that the ratio of old-population to B-type HVSs might be similar
to the ratio of the underlying populations near the GC.  This itself
is not known, but as proxy we evaluate the same ratio in the
solar neighborhood.  In fact, what is required is the ratio
of turnoff stars to B stars, since our proposed survey is
most sensitive to turnoff stars, while B stars form the only
population of HVSs that have been reliably tabulated.  The
distance ranges are similar: the B stars have typically been
detected at 60 kpc while turnoff stars at $g=21.5$ lie at about 25 kpc
(a factor 2.5 advantage for the B stars) but the B stars were searched
over 1 mag in $g$ while the turnoff stars would be searched over 2 mags
(which basically compensates).  

We estimate the ratio of $B$ stars to turnoff stars in the solar
neighborhood as follows.  We analyze samples of each stellar class drawn
from the Hipparcos catalog \citep{hip}, restricted to $V<7.3$
(Hipparcos completeness limit), distance $r<300\,$pc (to ensure
good parallaxes and low extinction), and distance from the Galactic plane
less than 50 pc.  We define turnoff stars as having absolute magnitudes
$3.5<M_V<4.5$ and near-turnoff colors ($0.3<B_T -V_T<0.8$).  For
B stars, we probe a 2-magnitude interval $-2.0<M_V<0.0$, which approximately
corresponds to the \citet{brown06} $g-r$ selection criterion, and we enforce
($-0.5<B_T -V_T<0.2$) to distinguish these from giants.  For each
class of star, we tabulate $\sum_i (V_{{\rm eff},i})^{-1}$,
where $V_{{\rm eff},i}$ is the ``effective volume'' over which that
star could have been found.  That is $V_{{\rm eff}}= 4\pi r_{\rm max}^3$
where $r_{\rm max}$ is the maximum distance that star could have been
detected given its observed absolute magnitude, its known direction,
and the two constraints on distance given above.  We find effective
densities of $0.91\times 10^{-5}\,\rm pc^{-3}$ and
$60.7\times 10^{-5}\,\rm pc^{-3}$ for the two classes, indicating
that turnoff stars are about 67 times more common that B stars.

Applying this rather crudely derived multiplier to the B-star HVS density
found by \citet{brown06}, we estimate that there could be 1 turnoff HVS
per $10\,\rm deg^2$.  Hence a survey of a few tens of square degrees
could probe the existence of this putative population.

\acknowledgments
We thank Alan Dressler and Mike Gladders for discussions
at OCIW morning tea that stimulated the genesis of this paper.
We are also grateful to Scott Gaudi, Jennifer Johnson,
Steve Shectman, and Ian Thompson, for helpful discussions.
J.A.K. was supported by NASA through a Hubble Fellowship.
A.G. was supported by grant AST 042758 from the NSF.
Any opinions, findings, and conclusions or recommendations expressed in
this material are those of the authors and do not necessarily reflect the
views of the NSF.

\end{document}